# An Effective EMTR-Based High-Impedance Fault Location Method for Transmission Lines

Jianwei An, Chijie Zhuang, *Member, IEEE,* Farhad Rachidi, *Fellow, IEEE* and Rong Zeng, *Senior Member, IEEE*

*Abstract*—This paper summarizes the electromagnetic time reversal (EMTR) technique for fault location, and further numerically validates its effectiveness when the fault impedance is negligible. In addition, a specific EMTR model considering the fault impedance is derived, and the correctness of the model derivation is verified by various calculation methods. Based on this, we found that when the fault impedance is large, the existing EMTR methods might fail to accurately locate the fault. We propose an EMTR method that improves the location effect of high-impedance faults by injecting double-ended signals simultaneously. Theoretical calculations show that this method can achieve accurate location for high-impedance faults. To further illustrate the effectiveness, the proposed method is compared with the existing EMTR methods and the most commonly used traveling wave-based method using wavelet transform. The simulation results show that the proposed double-ended EMTR method can effectively locate high-impedance faults, and it is more robust against synchronization errors compared to the traveling wave method. In addition, the proposed method does not require the knowledge or the a priori guess of the unknown fault impedance.

## I. INTRODUCTION

AS the scale of the power grid continues to expand, more and more high-voltage long-distance transmission lines are put into operation. Long-distance transmission lines are often faced with complex terrain, sometimes with harsh natural conditions, which makes long-distance power transmission more prone to failure. On the other hand, for long-distance transmission lines, the current inspection equipment in power system is not able to locate fault accurately and quickly, resulting in countless power system accidents. Therefore, locating faults in transmission networks is of great significance for the safety, stability and economic operation of the entire power grid.

Many transmission line fault location methods are available today [1]. In general, the existing fault location methods can be divided into two categories [2]: impedance-based (e.g., [3-5]) and traveling wave-based methods (e.g., [6-8]). In the impedance-based category, use is made of the power frequency signal to locate the line fault. Because of the relative simplicity of the method along with its reasonable cost in terms of the measuring equipment, it is widely used for fault location along transmission networks including high-impedance fault cases [9, 10]. However, the method is greatly affected by the fault impedance and the operating conditions prior to the fault. In contrast, the traveling wave-based methods use the transient traveling wave signal caused by the fault to determine the position of the fault point. Traveling wave-based methods are generally characterized by higher accuracy compared to impedance-based methods. Furthermore, they are not affected by factors such as the unknown fault impedance and line asymmetry. However, they require a relatively high frequency signal acquisition equipment, resulting in a higher cost compared to impedance-based methods. In addition, hybrid techniques that combine the above two methods to improve the effectiveness of fault location under complex configurations have also been proposed [11]. In general, the performance of current fault location methods has still room for improvement. Considering the increase of the reliability requirements of the grid power supply, further development is needed. Among the newly-proposed fault location methods in the past few years, electromagnetic time reversal (EMTR) is certainly one of the most interesting and promising methods to address the challenges of fault location in the future power grids.

The time reversal method was first applied to acoustics [12], and has been widely used in the correlation analysis of electromagnetic transient signals, such as radar communication [13], lightning location [14, 15], and electromagnetic wave focusing in biomedical applications [16]. In addition, EMTR-based transmission line fault location is also an important application. The application of the EMTR to the problem of fault location was first presented in [17], in which the terminal fault-generated transient voltage signal collected in the forward process is time reversed and re-injected into the network with the same topology using a Norton equivalent circuit. The short-circuit branch is set at different guessed positions along the line, and the position with the largest short-circuit current energy corresponds to the true fault position. Later, Razzaghi et al. presented a more comprehensive simulation and experimental verification of the method considering complex network structures such as branched and mixed (overhead and underground) networks[18, 19]. The effects of line losses as well as the use of different criteria to quantify the focusing of the time-reversed back-injected transients were discussed in [20-23]. Full-scale experimental validations are presented in An, Zhuang and Zeng are with State Key Lab of Power Systems, Department of Electrical Engineering, Tsinghua University, Beijing 100084, China.
Farhad Rachidi is with Electromagnetic Compatibility (EMC) Laboratory, Swiss Federal Institute of Technology Lausanne (EPFL), Switzerland.

[24, 25]. So far, most of the relevant research is based on the method proposed in [18]. Other EMTR-based methods, such as those proposed in [26-29] in which the medium in the backward-time does not match that in the forward time, have also been proposed. These methods do not require to carry out multiple backward-time simulations considering different guessed fault locations. For example, the method in [26, 27] allows to obtain the fault location by analysing the voltage phase along the line. On the other hand, Wang et al. [28, 29] have shown that the fault location can be determined by analysing the energy of the voltage signal in the backward-time phase.

In general, the existing research shows that the EMTR-based fault location methods are effective. At the same time, it should be noted that the current related research is not sufficient. In particular, the applicability of the EMTR-based methods to locate high-impedance faults has only been partially considered in [27, 29], and the effectiveness of the existing methods have not been quantified for locating high-impedance faults. High-impedance faults generally occur when an energized conductor gets connected to a surface with high resistance (such as a tree, dry ground, high-resistivity soil, or asphalt road, resulting in fault resistances of 100 Ω or more). Within this context, the focus of this paper will be the analysis of the EMTR-based methods to locate high-impedance faults.

The rest of this paper is organized as follows: Section II summarizes two existing EMTR-based fault location methods selected in the present study. In Section III, specific formulas for the existing single-ended EMTR fault location methods considering the fault impedance are derived, showing that they fail to locate high-impedance faults accurately. In Section IV, a double-ended EMTR method for either high or low impedance fault location is proposed, and its effectiveness is demonstrated by numerical simulations. Finally, general conclusions are given in Section V.

## II. CONSIDERED EMTR-BASED FAULT LOCATION METHODS

Although there are different EMTR-based methods for locating faults along transmission lines, we will consider in this paper two methods: (i) the classical method of Razzaghi et al. [18], and (ii) the method of Wang et al. [28]. These two methods are briefly summarized in the next two subsections.

### A. Razzaghi et al. Method [18] (EMTR-I)

As described in [18], the classical EMTR fault location process can be represented by Fig. 1 and Fig. 2. Consider a single-wire line above the ground terminated by known impedances. A fault occurs at a point $x_f$ along the line and is represented as an equivalent voltage source $U_f$. The resulting voltage signal at one of the terminals (such as $Z_0$) is measured in the forward time (Fig. 1). The signal is time-reversed and back-injected into the line using a Norton equivalent circuit (Fig. 2). The unknown short-circuit branch is varied along the line at different guessed fault locations $x'_f$ and the energy of the short-circuit are evaluated. In theory, it should be expected that the position with the highest energy corresponds to the fault location, i.e. $x'_f = x_f$.

The processes involved can be equivalently expressed in the frequency domain. Assuming that the voltage source applied to $x_f$ is $U_f(w)$, after the voltage signal is injected at the fault point, the voltage signal measured at the terminal load $Z_0$ is

$$U_0(\omega) = \frac{(1+\rho_0)e^{-\gamma x_f}}{1+\rho_0 e^{-2\gamma x_f}} U_f(\omega) \qquad (1)$$

where $\gamma$ is the line propagation constant. Since we assume that the line is lossless, $\gamma = j\beta$, with $\beta = \frac{2\pi f}{c}$ ($c$ denotes the propagation speed, and $f$ is the frequency). $\rho_0$ is

$$\rho_0 = \frac{Z_0 - Z_C}{Z_0 + Z_C} \qquad (2)$$

where $Z_C$ is the line characteristic impedance and $Z_0$ is the terminal impedance.

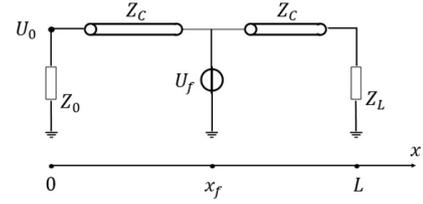

Fig. 1 Simplified representation of a fault along a transmission line. The fault is represented by an equivalent voltage source $U_f$

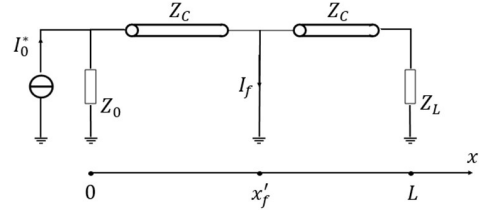

Fig. 2 Back-injection of the fault-generated transient signal using a Norton equivalent. The guessed fault location is at $x'_f$. (EMTR-I)

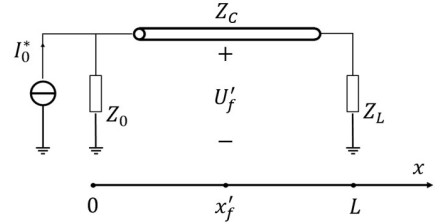

Fig. 3 Back-injection of the fault-generated transient signal using a Norton equivalent. The transverse branch is removed in the time-reversed phase. (EMTR-II)

The signal is reversed in time, equivalently complex conjugated in frequency, and re-injected into the network as shown in Fig. 2. The short-circuit current $I'_f$ at a guessed fault location $x'_f$ is

$$\begin{aligned} I'_f(x'_f, \omega) &= \frac{(1+\rho_0)e^{-\gamma x'_f}}{1+\rho_0 e^{-2\gamma x'_f}} I^*_0(\omega) \\ &= \frac{(1+\rho_0)e^{-\gamma x'_f}}{1+\rho_0 e^{-2\gamma x'_f}} \frac{U^*_0(\omega)}{Z_0} \\ &= \frac{(1+\rho_0)^2 \cdot e^{-\gamma(x'_f - x_f)}}{Z_0 \cdot (1+\rho_0 e^{-2\gamma x'_f}) \cdot (1+\rho_0 e^{2\gamma x_f})} U^*_f(\omega) \end{aligned} \qquad (3)$$

The energy of $I'_f$ is calculated using the Parseval's theorem. It is shown that the position where the signal energy of $I'_f$ is the largest satisfies $x'_f = x_f$, which can be used as a criterion for fault location.

*B. Mirrored Minimum Energy Method of Wang et al. [28] (EMTR-II)*

The forward-time step of the EMTR method proposed in [28] is the same as in the classical method of Razzaghi et al. [18], namely the recording of the fault-generated transient signal at one end of the transmission line. In the time-reversed step, the terminal voltage signal collected in the forward-time is time-reversed and back-injected using a Norton equivalent source into the same circuit, but in which the fault branch is removed (see Fig. 3). As opposed to the classical EMTR approaches, this method does not require to run multiple simulations to find the fault location.

The description of the entire process can also be derived in the frequency domain. In the forward-time step, the fault-generated voltage signal at the left terminal can be expressed using Eq. (1). In the time-reversed step, since the short-circuit branch is removed, the multiple reflection process of the entire line is considered. The voltage signal $U'_f$ is as follows:

$$U'_f(x'_f, \omega) = (1-\rho_0) \frac{e^{-\gamma x'_f} + \rho_L e^{-\gamma(2L-x'_f)}}{2 \cdot (1-\rho_0 \rho_L e^{-2\gamma L})} \cdot U_0^*(\omega)$$

$$= \frac{(1-\rho_0^2) \cdot (e^{-\gamma x'_f} + \rho_L e^{-\gamma(2L-x'_f)}) \cdot e^{\gamma x_f}}{2 \cdot (1-\rho_0 \rho_L e^{-2\gamma L}) \cdot (1+\rho_0 e^{2\gamma x_f})} \cdot U_f^*(\omega) \quad (4)$$

Similarly, the energy of the $U'_f$ signal can be calculated using the Parseval's theorem. Wang et al. [28] have shown that the minimum of the voltage signal energy computed in the reversed time occurs at the mirror image point of the fault location, with reference to the line centre.

*C. Verification of the effectiveness of the two considered EMTR-based fault location methods*

We consider a single-conductor transmission line characterized by typical parameters of power distribution networks, with a length of $L = 20$ km. The line is characterized by terminal impedances $Z_0 = Z_L = 100$ kΩ, and we assume that the fault signal is given by

$$U_f(\omega) = \frac{1}{j\omega} \text{V/(rad/s)} \quad (5)$$

By changing $x_f$ and $x'_f$ from 0 to $L$, we can calculate $I'_f$ and $U'_f$ using (3) and (4). The calculation results of the two EMTR-based methods are shown respectively in Fig. 4 and Fig. 5. As described above, these two methods allow to locate the fault location analysing the signal energy.

## III. PERFORMANCE OF EXISTING EMTR FAULT LOCATION METHODS FOR HIGH-IMPEDANCE FAULTS

In the derivations presented in Section II for the considered two EMTR-based models, the real fault impedance was not considered. Since we would like to assess the performance of the two EMTR fault location methods under different fault impedances, we need to derive an EMTR model that takes the fault impedance into account.

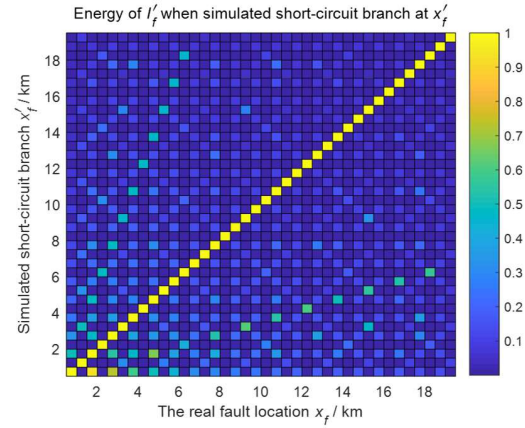

Fig. 4 Verification of the classical EMTR method (EMTR-I) of Razzaghi et al. [18]. For each true fault point position $x_f$, the energy value of the short-circuit current under different assumed fault branches can be calculated. The energy value at the diagonal position from the lower left corner to the upper right corner is the largest in each row, which indicates that the short-circuit branch current energy is always maximal at the true fault point for low-impedance faults.

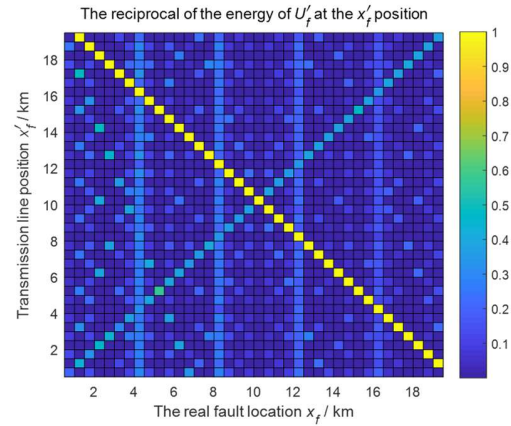

Fig. 5 Verification of the EMTR-based method (EMTR-II) of Wang et al. [28]. For each true fault point position $x_f$, the energy value of the voltage signal at different locations along the line can be calculated. The minimum of the voltage signal energy occurs at the mirror image point of the fault location, with respect to the line centre, for low-impedance faults.

*A. The left-end voltage $U_0$ collected in the forward-time process*

When considering the fault impedance $Z_f$, the circuit model for the forward-time process is shown in Fig. 6, and we can simplify it using Thévenin transformations [30], as shown in Fig. 7. First, the transmission line on the right-hand side of the fault point (Fig. 7a) can be replaced by the input impedance $Z_{in2}$, which is given by

$$Z_{in2} = Z_C \frac{1+\rho_L e^{-2\gamma(L-x_f)}}{1-\rho_L e^{-2\gamma(L-x_f)}} \quad (6)$$

Then, we further combine the fault branch and the line input impedance into a Thévenin equivalent circuit (Fig. 7b) with the following parameters

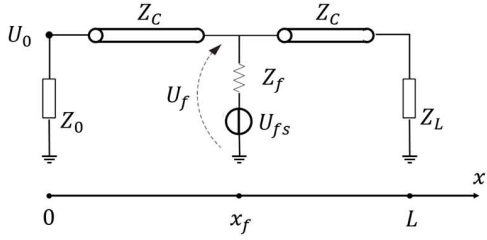

Fig. 6 Simplified representation of a non-zero impedance fault.

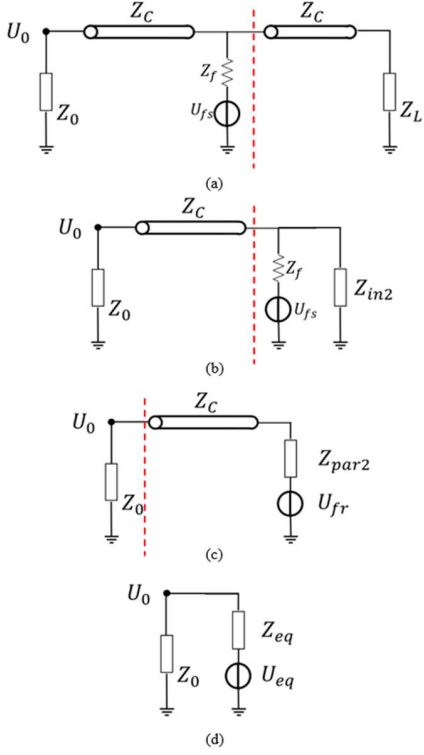

Fig. 7 Calculation of the left-end voltage $U_0$ by collapsing the network to its Thévenin equivalent circuit

$$Z_{par2} = \frac{Z_f Z_{in2}}{Z_f + Z_{in2}} \quad (7)$$

$$U_{fr} = \frac{Z_{in2}}{Z_{in2} + Z_f} U_{fs} \quad (8)$$

Next, we represent all the circuit to the right of the left termination using its Thévenin equivalent circuit as shown in Fig. 7(c), obtaining

$$Z_{eq} = Z_C \frac{1 + \rho_f e^{-2\gamma x_f}}{1 - \rho_f e^{-2\gamma x_f}} \quad (9)$$

$$U_{eq} = \frac{e^{-\gamma x_f}(1 - \rho_f)}{1 - \rho_f e^{-2\gamma x_f}} U_{fr} \quad (10)$$

where

$$\rho_f = \frac{Z_{par2} - Z_C}{Z_{par2} + Z_C} \quad (11)$$

The resulting equivalent circuit is shown in Fig. 7(d), from which $U_0$ can be obtained as

$$U_0 = \frac{Z_0}{Z_0 + Z_{eq}} U_{eq} \quad (12)$$

An analytical expression for the signal $U_0$ considering the fault impedance has been presented in [27]. We wish to emphasise that the analytical solution of [27] is somewhat different from the one of this paper. The difference will be explained in what follows.

The line input impedances at both sides of the fault point are $Z_{in1}$ and $Z_{in2}$ in Eq. (13) and Eq. (6):

$$Z_{in1} = Z_C \frac{1 + \rho_0 e^{-2\gamma x_f}}{1 - \rho_0 e^{-2\gamma x_f}} \quad (13)$$

In [27], after considering the fault source $U_{fs}$ voltage division, the voltage signal $U_f$ at the fault point of the line was derived as

$$U_f = \frac{Z_{par}}{Z_f + Z_{par}} U_{fs} \quad (14)$$

where

$$Z_{par} = \frac{Z_{in1} Z_{in2}}{Z_{in1} + Z_{in2}} \quad (15)$$

Then, [27] again represents the faulty branch and the right-hand side transmission line by an equivalent circuit. Considering the multiple reflection of $U_f$ between [0, $x_f$], the signal $U_0$ in [27] was obtained as follows:

$$U_0 = \frac{(1 + \rho_0) e^{-\gamma x_f}}{1 - \rho_f \rho_0 e^{-2\gamma x_f}} U_f \quad (16)$$

where $\rho_f$ is shown in Eq. (11).

However, it should be noted that when considering the input impedance of the left and right transmission lines in Eq. (14), multiple reflections of the signals transmitted to both ends have already been considered, therefore the multiple reflections on $U_f$ between [0, $x_f$] are duplicated in Eq. (16). In order to correct Eq. (16), the expression for the voltage $U_{fs}$ should be changed to

$$U_f = \frac{Z_{parL}}{Z_f + Z_{parL}} U_{fs} \quad (17)$$

with

$$Z_{parL} = \frac{Z_C Z_{in2}}{Z_C + Z_{in2}} \quad (18)$$

To validate Eq. (12), we took as reference simulation results obtained using ATP-EMTP software. The line is assumed to be 20 km long, the fault position is at 7 km, and the fault impedance is 100Ω. To represent the transfer function of the whole process, $U_{fs}$ was considered as a Dirac impulse. As shown in Fig. 8, the calculation results of Eq. (12) are completely consistent with those of ATP-EMTP. However, the results of the article [27] deviate from ATP-EMTP. In the same

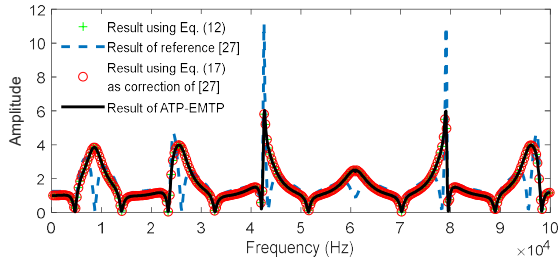

Fig. 8 Amplitude spectrum comparison of the signal $U_0$

Fig. 8, we have also added the results obtained using Eq. (16) in which $U_f$ is corrected using Eqs. (17) and (18), which are in perfect agreement with ATP-EMTP.

After the signal $U_0$ is calculated, the signals used for EMTR-based fault location can be obtained from either Eq. (3) (EMTR-I) or Eq. (4) (EMTR-II). Note in real applications of the EMTR-based fault location methods, the signal $U_0$ is not calculated, but recorded. We derive this formula only to obtained the transient signals that are required in the backward-time process.

### B. Performance of Existing EMTR Fault Location Methods for High-Impedance Faults

The same line described in Section II was used to test the fault location accuracy of the considered EMTR-based methods. We set different fault impedances, collected the transient signals, and then used the two EMTR methods described earlier to locate the fault and compare their location accuracy.

It should be emphasized that during the backward-time process, EMTR-I sets the value of the fault impedance to zero, since in practical cases this value is unknown. On the other hand, EMTR-II does not require the knowledge of the fault impedance (refer to Section II for details). The fault impedances are only used to calculate the transient signals in the forward-time step.

Fig. 9 presents the performance of the EMTR-I fault location method considering different fault resistances. It can be seen that when $Z_f = 1\Omega$, the calculated results are very similar to those of Fig. 4. However, the location accuracy of this method significantly degrades as $Z_f$ gradually increases. Beyond a fault resistance of $100\Omega$, the method fails in accurately locating the fault. Similarly, the calculated results for EMTR-II are shown in Fig. 10. When $Z_f = 1\Omega$, the calculated results are very similar to those shown in Fig. 5. As $Z_f$ gradually increases, the location accuracy of this method is also significantly deteriorated.

In summary, the existing two considered EMTR-based fault location methods can hardly achieve accurate fault location when the fault impedance exceeds a relative high value, e.g., 100 Ω.

### IV. A DOUBLE-ENDED EMTR-BASED METHOD FOR LOCATING HIGH-IMPEDANCE FAULTS

#### A. Proposed double-ended EMTR-based method (EMTR-III)

In the previous section, we have seen that the existing single-ended EMTR-based fault location methods are characterized by a poor location accuracy for high impedance faults. In this section, we will propose an EMTR-based method capable of accurately locating high-impedance faults.

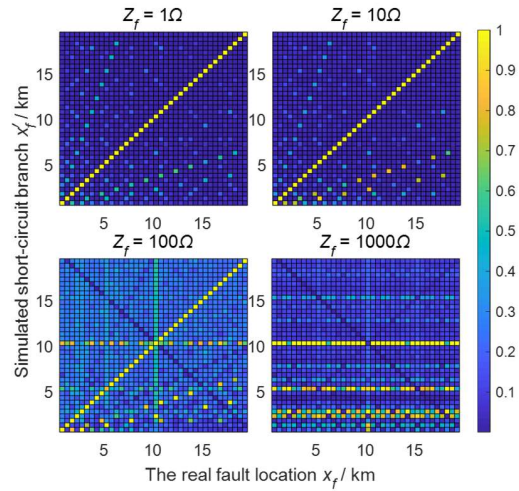

Fig. 9 Performance of EMTR-I fault location method under different fault impedances. When the fault impedance increases to values above 100 Ω, the predicted fault locations deviate from the real ones (for these cases, the position with the highest energy in the figure is not in the diagonal position from the lower left corner to the upper right corner).

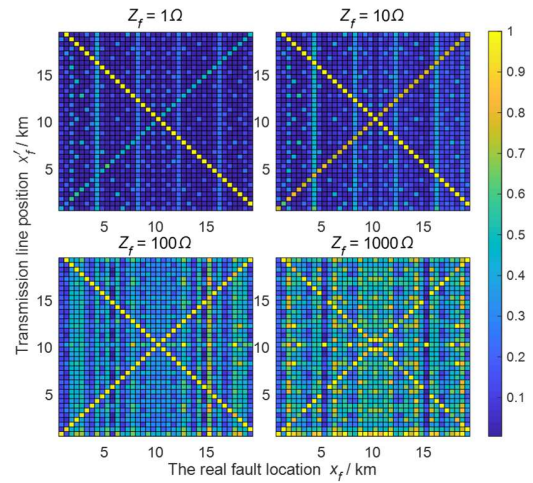

Fig. 10 Performance of EMTR-II fault location method under different fault impedances. When the fault impedance increases to values above 100 Ω, the predicted fault locations deviate from the mirror position of the real ones (for these cases, the position with the highest energy in the figure is not in the diagonal position from the upper left corner to the lower right corner).

We first explain why the existing EMTR method is effective when the fault impedance is small. When the forward process contains a short-circuit branch, the signal measured at $Z_0$ is always reflected back and forth between $[0, x_f]$. When the signals collected in the forward process is time-reversed and injected back into the line of Fig. 2, when the guessed fault location coincides with the real one, $x'_f = x_f$, the wave injected by the current source and the wave reflected by the short−circuit point completely overlap, resulting in a maximal short−circuit energy at the short−circuit point. However, when the fault impedance is large, the signal is not only reflected between $[0, x_f]$, but will be refracted to the right of the fault point, resulting in other signal components. This makes the wave

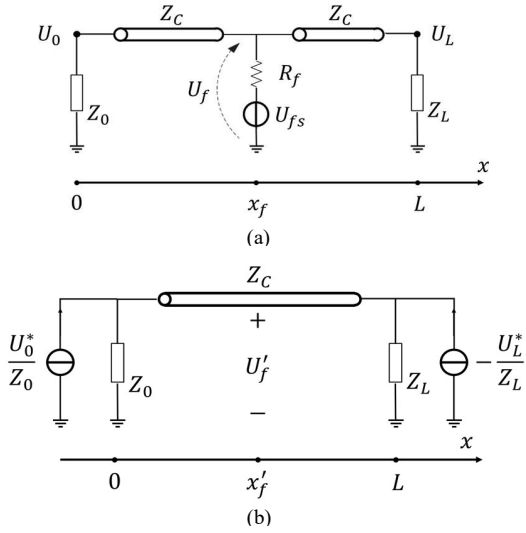

Fig. 11 Representation of the proposed EMTR-based double-end method (EMTR-III). (a) Forward-time process, and (b) backward-time process. Note that $U_L$ is time-reversed and multiplied by -1.

Table 1 Algorithm for the double-ended EMTR-based method

| |
|---|
| **Input:** network parameters $(L, Z_C, \gamma, Z_0, Z_L)$ and voltage signals $U_0$ and $U_L$ collected at both ends of the line after a fault occurs. |
| 1. Both $U_0$ and $U_L$ signals are reversed in the time domain (complex conjugate is taken in the frequency domain), and $U_L$ is multiplied by -1.
2. The signals (after step 1) at both ends are reinjected into the network, after having removed the fault branch.
3. For each location $x'_f$ on the line:
   Calculate the energy of the voltage signal at $x'_f$.
4. Extract the location $x'_{fmin}$ corresponding to the minimum voltage signal energy. |
| **Output:** the predicted fault location is $x_f = L - x'_{fmin}$. |

injected by the current source and the wave reflected by the short−circuit point no longer completely overlap when $x'_f = x_f$ in the time-reversed step, thus reducing the ability of locating the fault point.

In order to cope with the above-mentioned problem and achieve a better location accuracy as the fault impedance increases, we propose to introduce a second measurement station and consider a double-ended fault location method in which, the fault-generated signals at both ends are time-reversed, and synchronously back-injected into the circuit, in absence of the fault branch. The basic idea here is to eliminate the signal refracted beyond the fault point by introducing double-ended signals, with the signal at the second end time-reversed and multiplied by -1, thereby achieving a location accuracy similar to that of a low-impedance short circuit. The specific algorithm of this method is illustrated in Fig. 11 and described in Tab. 1. Note that the additive inverse operation to $U_L$ (multiplication by -1) is necessary to satisfy the mirrored minimum energy criterion.

The voltage expressed in the frequency domain at an arbitrary location $x'_f$ along the line as a result of the back-injection of time-reversed signals at both ends can be readily obtained as follows:

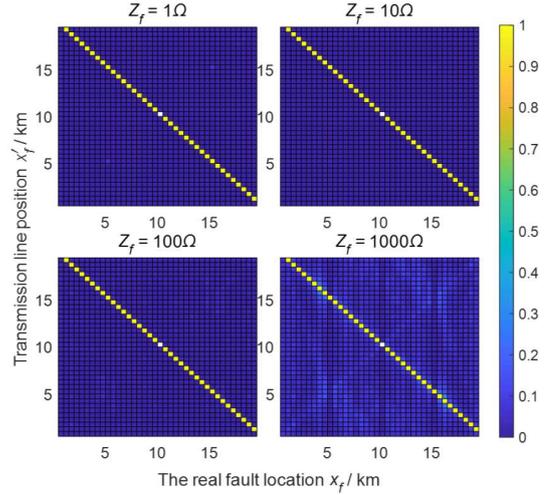

Fig. 12 Performance of the proposed EMTR-III fault location method under different fault impedances. For each true fault point position $x_f$, the energy value of the voltage signal at different locations along the line can be calculated. It can be seen that the minimum of the voltage signal energy occurs at the mirror image point of the fault location, with respect to the line centre. The results show that accurate positioning can still be achieved even when the fault impedance is large.

$$U_f(x'_f, \omega) = \frac{(1-\rho_0)\cdot\left(e^{-\gamma x'_f} + \rho_L e^{-\gamma(2L-x'_f)}\right)}{2\cdot(1-\rho_0\rho_L e^{-2\gamma L})} \cdot U_0^*$$
$$-\frac{(1-\rho_L)\cdot\left(e^{-\gamma(L-x'_f)} + \rho_0 e^{-\gamma(L+x'_f)}\right)}{2\cdot(1-\rho_0\rho_L e^{-2\gamma L})} \cdot U_L^* \quad (19)$$

where $U_0$ is given in Eq. (12), while an analytical expression for $U_L$ can be readily obtained, in a similar manner to $U_0$.

As a preliminary validation, we considered the transmission line parameters in Section II(c) to test the method under different impedances. The results are shown in Fig. 12. Compared with Fig. 9 and Fig. 10, the double-ended EMTR can achieve accurate positioning of the fault even when the fault impedance is as high as $Z_f = 1000\ \Omega$.

### B. Location accuracy assessment of the EMTR double-ended method (EMTR-III)

In this section, we test the three EMTR fault location methods and the commonly used traveling wave location method based on the wavelet transform (WT)[7, 8] to locate the faults with different impedances, and compare the location accuracy of each method. Wavelet transform is one of the most used feature extraction methods for various fault diagnosis systems [31, 32]. Using WT, the wavefront arrival time of the wavefront, and thus the time difference of arrival, can be precisely detected, allowing an accurate location of the fault.

The schematic diagram of the single-phase transmission line used in the simulation is shown in Fig. 13. A constant-parameter model is used to model the transmission line, which is the same as the [24, 28]. The parameters of transmission line are given in Table 2. The line is assumed to be terminated at its both ends on power transformers represented by high impedances, namely 100 kΩ. The power supply of the line is provided by an AC voltage source placed at $x = 0$. To generate precise signals, the simulation time step was set to 0.1 µs and then sampled at 1 MHz, the same as many existing traveling wave-based fault location methods [33, 34]. In addition, it is assumed that the signal-to-noise ratio of the signal acquired in the forward-time process is 30 dB. The calculation results are shown in Table 3 and Table 4, considering different fault location points (20, 40, 60 and 80 km), as well different impedances (50 and 300 Ohms).

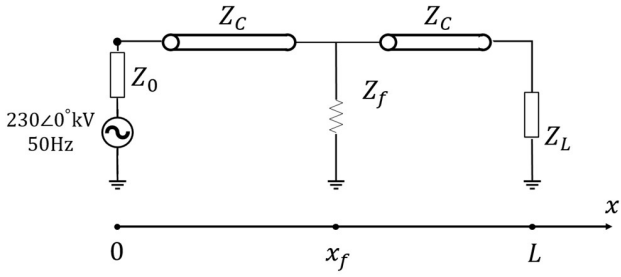

Fig. 13 Schematic representation of the transmission line.

Table 2 Adopted transmission line parameters.

| Line Parameter | Value |
| --- | --- |
| Length | 100 km |
| Per-unit length inductance | 1.60 µH/m |
| Per-unit length capacitance | 10.54 pF/m |
| Per-unit length resistance | 0.036 mΩ/m |

Table 3 Fault location errors (fault impedance is 50Ω)

| Fault location / km | 20 | 40 | 60 | 80 |
| --- | --- | --- | --- | --- |
| EMTR-I | 0.021 | 0.044 | 0.08 | 0.04 |
| EMTR-II | 0.09 | 0.06 | 0.1 | 0.06 |
| EMTR-III | 0.012 | 0.014 | 0.015 | 0.012 |
| WT | 0.0975 | 0.0971 | 0.0726 | 0.0733 |

Table 4 Fault location errors (fault impedance is 300Ω)

| Fault location / km | 20 | 40 | 60 | 80 |
| --- | --- | --- | --- | --- |
| EMTR-I | 30 | 10 | 10 | 30 |
| EMTR-II | 6 | 0.6 | 0.4 | 0.4 |
| EMTR-III | 0.04 | 0.073 | 0.074 | 0.05 |
| WT | 0.1372 | 0.1108 | 0.1003 | 0.113 |

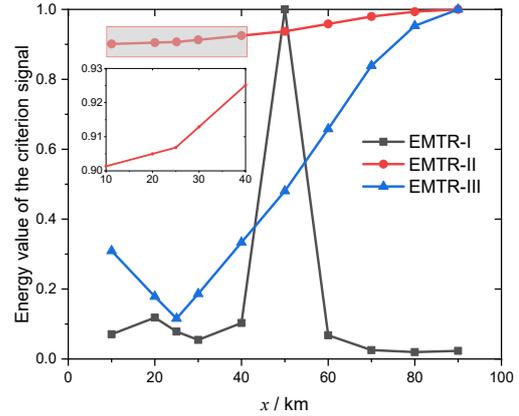

Fig. 14 The energy of the voltage signal along the line for the three considered EMTR-based methods. The true fault location is at x = 75 km. Neither EMTR-I nor EMTR-II was able to locate the fault, while EMTR-III predicts the minimum of the voltage signal energy occurring at the mirror image point of the fault location with respect to the line center, namely at x = 25 km.

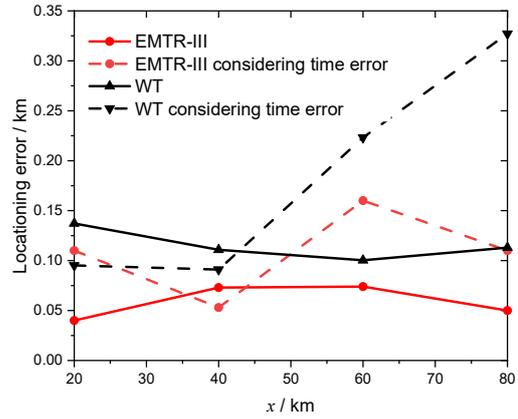

Fig. 15 Location error considering a time synchronization error of 1µs

It can be seen that the location error of the four considered fault location methods is below 100 m when the fault impedance is 50Ω, and EMTR-III, the method proposed in this paper, is characterized by the highest accuracy (location error below 15 m). For the case of high fault impedance (300Ω), fault location methods EMTR-I and EMTR-II fail in accurately locating the fault points. On the other hand, EMTR-III and WT methods still achieve accurate fault location results. Moreover, the accuracy of EMTR-III is higher than WT. It should be noted that although EMTR-II method has a good location accuracy at some fault locations when the fault impedance is large, as can be seen from Fig. 10, but in general, the location cannot be predicted reliably. An example is shown in Fig. 14 where the fault point is at 75 km, while both EMTR-I and EMTR-II fail to locate the fault, EMTR-III can accurately locate it.

### C. *Effect of time error on fault location accuracy*

EMTR-III method proposed in this paper requires to record the fault-generated signals at both ends of the line in a time-synchronized manner. Theoretically, since the EMTR method relies on the temporal and spatial relationship between multiple reflected signals, it may be less sensitive to the synchronization error than the wavelet-based method is, because the wavelet-based method relies essentially on the initial portion of the observation point signal.

By shifting the signal at one terminal of the line, we can set the synchronization error. We conducted the same simulations using the two methods assuming a synchronization error of 1µs. The resulting location errors are shown in Fig. 15 for both methods. It can be seen that EMTR-III method is robust against such synchronization errors and performs better than WT method.

V. CONCLUSIONS

In this paper, we discussed the performance of two existing EMTR-based fault location methods, when applied to locate high-impedance faults. We showed that when the fault impedance is large, the existing EMTR methods are not able to reliably locate the position of the fault.

To increase the efficiency of EMTR-based method to locate high-impedance faults, we proposed a double-ended method.

The proposed method was found to be effective in locating faults with impedances as high as 1000 Ohms. The simulation results showed also that the proposed method is robust against synchronization errors and performs better than a traditional wavelet-based fault location method.

Last but not least, the proposed method does not require the knowledge of the unknown fault impedance, since in the time-reversed step, the fault branch is removed from the network.

We wish to remark that the analysis of the method proposed in this paper does not consider nonlinear arc models, which will be the subject of our future work. In addition, a rigorous mathematical proof for the proposed method has not been presented in the paper. Work is in progress to achieve this task.

**Acknowledgments** - This work is partly supported by National Key Research and Development Program of China under Project No. 2017YFB0902701, National Natural Science Foundation of China under grants 51921005, State Grid Corporation of China under project 52120519000M, Sichuan Energy Internet Research Institute of Tsinghua University, and Swiss Federal Office of Energy (Contract SI/501706-01).